\documentclass[aps,twocolumn,superscriptaddress,showpacs]{revtex4}
\usepackage{epsfig,graphicx}
\usepackage{indentfirst}
\usepackage{amsfonts,amssymb,latexsym,amsmath,enumerate,amsthm}
\usepackage{bm}
\usepackage{color}

\newcommand{\be}{\begin{equation}}
\newcommand{\ee}{\end{equation}}
\newcommand{\bea}{\begin{eqnarray}}
\newcommand{\eea}{\end{eqnarray}}

\begin{document}

\title{Gaussian and Airy wave-packets of massive particles\\ with orbital angular momentum}

%
\author{Dmitry~V.~Karlovets}
    \email[E-mail: ]{d.karlovets@gmail.com}
\affiliation{Max-Planck-Institut f\"ur Kernphysik, Saupfercheckweg 1, D-69117 Heidelberg, Germany}
\affiliation{Tomsk Polytechnic University, Lenina 30, 634050 Tomsk, Russia}
%

\date{\today}

\begin{abstract}
While wave-packet solutions for relativistic wave equations are oftentimes thought to be approximate (paraxial), we demonstrate that there is a family of such solutions, which are exact, by employing a null-plane (light-cone) variables formalism. A scalar Gaussian wave-packet in transverse plane is generalized so that it acquires a well-defined z-component of the orbital angular momentum (OAM), while may not acquire a typical ``doughnut'' spatial profile. Such quantum states and beams, in contrast to the Bessel ones, may have an azimuthal-angle-dependent probability density and finite uncertainty of the OAM, which is determined by the packet's width. We construct a well-normalized Airy wave-packet, which can be interpreted as a one-particle state for relativistic massive boson, show that its center moves along the same quasi-classical straight path and, what is more important, spreads with time and distance exactly as a Gaussian wave-packet does, in accordance with the uncertainty principle. It is explained that this fact does not contradict the well-known ``non-spreading'' feature of the Airy beams. While the effective OAM for such states is zero, its uncertainty (or the beam's OAM bandwidth) is found to be finite, and it depends on the packet's parameters. A link between exact solutions for the Klein-Gordon equation in the null-plane-variables formalism and the approximate ones in the usual approach is indicated, generalizations of these states for a boson in external field of a plane electromagnetic wave are also presented.
\end{abstract}


\maketitle

\section{Introduction}

As was demonstrated in 1990s first theoretically and then experimentally, laser beams and even single photons with a doughnut spatial profile can carry orbital angular momentum (OAM) quantized along the propagation axis (see e.g.,\cite{Mono} and references therein). Appearance of this orbital angular momentum owes to the peculiar (helical) spatial structure of the photonic beam and not to the polarization degree of freedom. Some 15 years later, in 2007, somewhat more intricate photon beams were brought into life, namely the Airy beams having such features as being spreading free, self-healing and moving along a curvilinear trajectory without an external force \cite{Airy_Exp}. More recently, in 2010-2011, several groups managed to create massive particles, namely electrons with a kinetic energy of $\sim 300$ keV, carrying orbital angular momentum with its quanta up to $100\hbar$ \cite{Uchida, Verbeeck, McMorran, Uchida_2012}. Finally, the electronic Airy beams of the energy of $\sim 200$ keV  were also created experimentally just recently \cite{Airy_El_Exp}. All these novel quantum states open up many different perspectives in quantum optics \cite{Mono}, in electron microscopy and material properties studies \cite{Verbeeck}, in physics of electromagnetic radiation \cite{We}, and even in the high-energy physics \cite{Serbo, Ivanov, I-S, Ivanov_PRD}.

Theoretical studies of these non-plane-wave states (or beams) commonly deal with the ``pure Bessel'' or the ``pure Airy'' states having the features of being non-normalizable in infinite volume, exactly as the plane waves (see e.g., \cite{Bliokh-11, Airy}). When applying these simplified states to the real physical problems, one has to either quantize them in a finite volume \cite{Ivanov} or add some (usually, Gaussian) envelope function, thus turning these states into the wave-packets \cite{Airy_beam, I-S}. This is where the theoretical studies for massive bosons and fermions are not so vast yet. First of all, the spreading feature is inherent to all the feasible quantum wave packets, since it is closely connected with the coordinate-momentum uncertainty relations. While introduction of an envelope makes the energy of an Airy beam finite \cite{Airy_beam} (unlike in the ``pure Airy'' case), this simultaneously brings about necessity of spreading -- the feature that is known to absent for Airy beams. This contradiction is worth exploring in more detail.

Secondly, the wave-packet states used for describing the observed electrons and photons represent approximate (paraxial) solutions of the corresponding wave equations, whereas when applying these states, for instance, to the quantum scattering problems in the high-energy physics it is highly desirable to have appropriate \textit{exact solutions}. Here, we show that such solutions for the Klein-Gordon equation can be obtained in the well-known formalism of the null-plane (light-cone) variables. A new set of exact solutions for relativistic wave equations represents an independent interest also for mathematical physics, of course.

Thirdly, as we know from optics with these twisted (or vortex) photons, practical interest oftentimes represent even not the simple Gaussian-Bessel wave-packets with azimuthally-symmetrical intensity profile \cite{I-S} but somewhat more sophisticated superpositions of the OAM eigenstates with the 2D Gaussian packets \cite{Mair_2001}. For instance, photonic states having finite quantum uncertainty of the OAM (or the beam's OAM bandwidth, see e.g.\cite{Torres_2003, F-A_EPJD_2012}) were used for creating pairs of photons entangled in their OAM values after the parametric down-conversion process, and the feature of non-vanishing OAM bandwidth turned out to be crucially important \cite{Mair_2001, F-A_2012}. To the best of our knowledge, exact wave-packet states of this type for massive particles have not been presented before, which, in particular, hampers development of the idea to create OAM-entangled pairs of electrons, protons and other massive particles \cite{Ivanov_PRA}.  

Finally, the quantum wave-packet states with such degrees of freedom as, for instance, OAM can exist even in some external electromagnetic fields, if operators associated with the corresponding quantum numbers commute with the (Klein-Gordon or Dirac) Hamiltonian \cite{Non-Volkov}. A plane electromagnetic wave counter-propagating to the particle is known to leave the particle's effective transverse dynamics unchanged. Physically, it means that if a twisted particle is brought into collision with a laser pulse, effective OAM of the former, unlike the linear momentum, would not change considerably when leaving the wave (neglecting the radiation losses). Corresponding solutions for relativistic wave equations with an external field could allow one to calculate radiation (scattering) processes with the particles in these new quantum states non-perturbatively (in the Furry picture \cite{BLP}).

In this paper, we study in detail wave-packets carrying OAM and generalize the well-known OAM-less quasi-classical Gaussian states (the so-called squeezed partially coherent states \cite{B-G, Unbound, Unbound_2}). Mathematically, the wave-packets being described belong to one class of exact solutions for Klein-Gordon equation having quasi-classical averages (in initial moment of time) and differing from each other in momentum representation only in the general complex phase. We demonstrate, in particular, that there are some states (and beams), which have a well-defined z-projection of the OAM but do not have a typical ``doughnut'' spatial profile of the probability density. This extends validity of the Berry's statement in optics that there is no direct relation between phase vortices and the OAM \cite{Berry}. A physical reason for absence of this azimuthally-symmetric profile is a finite quantum uncertainty of the orbital angular momentum, whose value is determined by the packet's width. This result is generalized for a boson in external field of the electromagnetic wave.

We also present well-normalized Airy wave-packet states of a boson (including these states in the external field) and calculate such averages as trajectory, dispersions and the OAM. It is shown that effective value of the Airy packet's OAM is zero, analogously to the optical case \cite{OAM-Airy}, and the spreading rate of the packet coincides with that of the ordinary Gaussian beam (i.e. in accordance with the uncertainty relations). We explain this feature by noting that this spreading occurs at the expense of the exponential ``tail'', which only makes the Airy packet normalizable, while its central peak stays practically unchanged in time, in agreement with the experiments. At the same time, we found that quantum uncertainty of the OAM for these Airy wave-packets (or the beam's OAM bandwidth) is finite, and it is determined by the packet's parameters.  Another interesting observation here is that this quantum OAM-uncertainty does not turn into zero even in the case of the vanishing transverse momentum, unlike the one of the Gaussian beam. This means that while the OAM-bandwidth of the Gaussian beam may be said to be extrinsic (in terms of Ref.\cite{Neil}) because its value depends upon the choice of the quantization axis, the Airy beam's OAM-bandwidth has both extrinsic and \textit{intrinsic} contributions.

The feature of finiteness of the OAM uncertainty, both for Gaussian wave-packets and the Airy ones, could be rather useful in view of possible experiments with the OAM-entangled electrons and other massive particles. In quantum optics, it is this feature that determines degree of entanglement of the down converted photons, and the entanglement practically vanishes for the ``pure'' OAM states (see e.g.,\cite{F-A_2012}). 

The structure of the paper is as follows. In Sec. II we describe basic properties of the OAM-possessing quantum states, and in Sec. III we show how one can reduce the 4D Klein-Gordon equation to the easier 2D Schr\"odinger one by using the light-cone variables formalism. In Sec. IV we describe relativistic bosons' wave-packets without OAM, including Airy one-particle states, in momentum and coordinate representations. In Sec. V, we generalize them adding OAM quantized along the z-axis, calculate the OAM uncertainty and discuss some features as well as possible generalizations of the Airy states. Similar wave-packets in external field of a plane electromagnetic wave are presented in Sec. VI. A system of units $\hbar = c = 1$ is used throughout the paper.

\section{Quantum states with OAM}

A quantum-mechanical approach to the one-particle states with the OAM was developed earlier -- see e.g., Refs. \cite{Bliokh-11, me}. Here we employ quantum-field-theoretic methods, which are more convenient for the purposes of quantum scattering problems. In a general case, the scalar quantum states with the OAM, $|p_\parallel, \kappa, \ell \rangle$, represent complete and orthogonal set, which obeys the following orthogonality relation:
\begin{eqnarray}
&& \displaystyle \langle p_\parallel^{\prime}, \kappa^{\prime}, \ell^{\prime}|p_\parallel, \kappa, \ell \rangle = \cr && \qquad \displaystyle = (2\pi)^2 2 \varepsilon ({\bm p})\, \delta (p_{\parallel} - p_{\parallel}^{\prime}) \frac{\delta (\kappa - \kappa^{\prime})}{\kappa}\, \delta_{\ell \ell^{\prime}} \label{Eq2.5aa}
\end{eqnarray}
where: $\varepsilon ({\bm p}) = \sqrt{{\bm p}^2 + m^2} \equiv \sqrt{p_{\parallel}^2 + \kappa^2 + m^2}$, $\kappa$ is an absolute value of the transverse momentum, $p_\parallel$ is a longitudinal component of the momentum, $\ell = 0, \pm 1, \pm 2, ...$ is the OAM, and the momentum's azimuthal component remains unfixed. The one-particle states with the OAM in coordinate representation can be defined in the usual way:
\begin{eqnarray}
&& \displaystyle \psi_{\{p_{\parallel}, \kappa, \ell\}} (x) : = \frac{1}{\sqrt{2 \varepsilon}} \langle {\bm x}| p_{\parallel}, \kappa, \ell\rangle = \langle {\bm x}| \hat{a}^{\dagger}_{\{p_{\parallel}, \kappa, \ell\}}|0\rangle = \cr && \qquad \displaystyle = \int \frac{d^3 p}{(2\pi)^3 \sqrt{2 \varepsilon ({\bm p})}} \langle {\bm p}| p_{\parallel}, \kappa, \ell \rangle e^{-ipx} = \cr && \qquad \qquad \qquad \displaystyle = J_{\ell} (\kappa \rho) e^{-i\varepsilon t + ip_{\parallel}z + i \ell \phi_r}. \label{Eq2.5a}
\end{eqnarray}
Here we used $\langle {\bm x}|{\bm p}\rangle = \langle {\bm x}|\sqrt{2 \varepsilon({\bm p})}\ \hat{a}^{+}_{{\bm p}}|0\rangle = \sqrt{2 \varepsilon ({\bm p})} e^{-ipx}$ (see e.g., Ref. \cite{P-Sh}), and $J$ denotes a Bessel function. Hence, in momentum representation the one-particle state is:
\begin{eqnarray}
&& \displaystyle \langle {\bm p}|\hat{a}^{\dagger}_{\{p_{\parallel}, \kappa, \ell\}}|0\rangle = \cr && \qquad \displaystyle = (-i)^{\ell} (2\pi)^2 \delta (p_z - p_{\parallel}) \frac{\delta (p_{\perp} - \kappa)}{p_{\perp}}\, e^{i \ell \phi_{p}}. \label{Eq2.5}
\end{eqnarray}

Creation and annihilation operators commute as follows: 
\begin{eqnarray}
&& \displaystyle [\hat{a}_{\{p_{\parallel}, \kappa, \ell\}}, \hat{a}^{\dagger}_{\{p_{\parallel}^{\prime}, \kappa^{\prime}, \ell^{\prime}\}}] = \cr && \qquad \displaystyle = (2\pi)^2 \delta (p_{\parallel} - p_{\parallel}^{\prime})\frac{\delta (\kappa - \kappa^{\prime})}{\kappa}\, \delta_{\ell\ell^{\prime}},\label{Eq2.9}
\end{eqnarray}
and all the rest are zero. The secondary-quantized field operator can be written by analogy with the plane-wave case:
\begin{eqnarray}
&& \displaystyle \hat{\psi} (x) = \sum\limits_{\ell} \int \frac{dp_{\parallel} \kappa d\kappa}{(2\pi )^2 2 \varepsilon} \left (\langle {\bm x}| p_{\parallel}, \kappa, \ell \rangle\ \hat{a}_{\{p_{\parallel}, \kappa, \ell\}} + \text{h.c.}\right ) \cr && \displaystyle = \sum\limits_{\ell} \int \frac{dp_{\parallel} \kappa d\kappa}{(2\pi )^2 \sqrt{2 \varepsilon}} \Big (J_{\ell} (\kappa \rho) e^{-i\varepsilon t + ip_{\parallel}z + i \ell \phi_r} \hat{a}_{\{p_{\parallel}, \kappa, \ell\}} + \cr && \qquad \qquad \qquad \qquad \qquad \displaystyle + \text{h.c.}\Big ), \label{Eq2.10}
\end{eqnarray}
so that the one-particle state, as might be easily checked, will be $\psi_{\{p_{\parallel}, \kappa, \ell\}} (x) = \langle 0|\hat{\psi}(x)|p_{\parallel}, \kappa, \ell \rangle$. Generalization of these formulas for the vectorial or spinor fields is straightforward. Using Eq.(\ref{Eq2.9}), the commutator for field operators is found as:
\begin{eqnarray}
&& \displaystyle [\hat{\psi} (x), \hat{\psi}^{\dagger} (x^{\prime})] = \sum \limits_{\ell} \int \frac{dp_{\parallel}\kappa d\kappa}{(2\pi)^2 2\varepsilon} J_{\ell}(\kappa \rho)J_{\ell}(\kappa \rho^{\prime})\cr && \displaystyle \times \left(e^{-i\varepsilon (t-t^{\prime}) + ip_{\parallel}(z-z^{\prime}) + i\ell(\phi_r - \phi_r^{\prime})} - \text{c.c.}\right). \label{Eq2.12}
\end{eqnarray}
In fact, the r.h.s. here is a standard Pauli-Jordan function (see e.g., Ref. \cite{B-Sh}). Indeed, the summation over $\ell$ can be done using the formula (8.530) in Ref.\cite{Ryzh} and after that it is easy to show that the field commutator coincides with its commonly-used ``plane-wave'' form.

\section{Some Schr\"odinger non-plane-wave states}

The non-plane-wave quantum states for relativistic bosons and fermions (no matter free or the ones put in some background field) can be obtained with the use of at least two methods: 

In the first one, we reduce the differential equation under consideration (the Klein-Gordon or the Dirac one) to the somewhat easier Schr\"odinger equation, for which the set of known exact solutions is far more vast (see e.g., \cite{B-G, Non-Volkov}). This method is mathematically rigorous and elegant yet, at the same time, it lacks for explicit Lorentz invariance, and a physical interpretation of the Schr\"odinger equation's solutions might remain somewhat hidden. 

In the second approach, we represent a quantum state as a superposition of ones forming an orthonormal set of the known exact solutions for the corresponding equation and then choose the overlap of these states according to the desired physical model. This approach leaves more freedom, it preserves relativistic invariance (to a needed extent), and it seems to be more physically illustrative. At the same time, such properties of the resultant solutions as orthogonality and completeness are not obvious here, so they should be checked separately. In what follows, we will combine both approaches for complementarity.

Firstly, let us illustrate how the Klein-Gordon equation could be reduced to the 2D Schr\"odinger one (see e.g, \cite{B-G, Non-Volkov}), for which many non-plane-wave solutions are already known. The Klein-Gordon equation, 
$$
\hat{K} \Psi (x) = 0,\, \hat{K} = \hat{p}^2 - m^2,\ \hat{p}^{\mu} = i\partial^{\mu},
$$ 
written in terms of the null-plane (the light-cone) variables (see e.g., \cite{B-G, Rohrlich}), 
\begin{eqnarray}
& \displaystyle \xi : = (nx) = t + z,\ \tilde{\xi} : = (\tilde{n}x) = t - z,\cr & \displaystyle n = \{1, 0 , 0, -1\},\, \tilde{n} = \{1, 0 , 0, 1\},\, n^2  = \tilde{n}^2 = 0, \label{Sh.0}
\end{eqnarray}
has the following form
\begin{eqnarray}
& \displaystyle \left (4 \partial^2_{\xi \tilde{\xi}} + \hat{\bm p}_{\perp}^2 + m^2 \right ) \Psi (\xi, \tilde{\xi}, {\bm r}_{\perp}) = 0. \label{Sh.1}
\end{eqnarray}
A more general case with the vectors $n = \{1, {\bm n}\},\, \tilde{n} = \{1, -{\bm n}\},\, {\bm n}^2 = 1$ may be obtained by a simple rotation of the coordinates. Then if one considers a ``monochromatic'' state:
\begin{eqnarray}
& \displaystyle \hat {\lambda} \Psi (\xi, \tilde{\xi}, {\bm r}_{\perp}) = \lambda \Psi (\xi, \tilde{\xi}, {\bm r}_{\perp}),\cr & \displaystyle \hat{\lambda} = (n\hat{p}) = \hat{p}^0 + \hat{p}^3 = 2i \partial_{\tilde{\xi}}, \label{Sh.2}
\end{eqnarray}
we will have
\begin{eqnarray}
& \displaystyle \Psi (\xi, \tilde{\xi}, {\bm r}_{\perp}) = \psi (\xi, {\bm r}_{\perp}) \exp\left\{-\frac{i}{2} \lambda \tilde{\xi}\right\}. \label{Sh.3}
\end{eqnarray}
One can get rid of the mass term by substituting $\psi (\xi, {\bm r}_{\perp}) \propto \exp\{-i m^2 \xi/2\lambda\}$. Changing variables to the dimensionless ones,
\begin{eqnarray}
& \displaystyle \tau = 2 \lambda \xi,\, {\bm x}_{\perp} = 2 \lambda {\bm r}_{\perp},\, \text{so that}\cr & \displaystyle \Psi (\xi, \tilde{\xi}, {\bm r}_{\perp}) = \Phi (\tau, {\bm x}_{\perp}) \exp\left\{-\frac{i}{2} \lambda \tilde{\xi} -i \frac{m^2 \xi}{2\lambda}\right\},\label{Sh.4}
\end{eqnarray}
we arrive at the free particle's 2D Schr\"odinger equation:
\begin{eqnarray}
\displaystyle \left (i\partial_{\tau} - \hat{H}\right )\Phi (\tau, {\bm x}_{\perp}) = 0,\, \hat{H} = - \Delta_{\perp} = - \partial_x^2 - \partial_y^2, \label{Sh.5}
\end{eqnarray}
There are several non-plane-wave solutions to this equation. Here, we mention only the ones we will need in this paper.

Perhaps, the best known non-plane-wave solution is a normalized to unity (on a plane $\tau = \text{const}$) Gaussian wave-packet (or a squeezed partially coherent state -- see e.g., Refs. \cite{B-G, Unbound, Unbound_2}),
\begin{eqnarray}
&& \displaystyle \Phi (\tau, {\bm x}_{\perp}) = \sqrt{\frac{a}{\pi}} \frac{1}{a + 2i\tau} \exp \Big \{i\bar{\bm p}_{\perp} \left ({\bm x}_{\perp} - \frac{1}{2} \bar{\bm x}_{\perp}\right ) - \cr && \qquad \qquad \qquad \qquad \qquad \qquad \displaystyle - \frac{1}{2} \frac{\left ({\bm x}_{\perp} - \bar{\bm x}_{\perp}\right )^2}{a + i2\tau}\Big\}  \label{Sh.6}
\end{eqnarray}
which has the quasi-classical averages
\begin{eqnarray}
& \displaystyle \langle{\bm x}_{\perp}\rangle = \bar{\bm x}_{\perp} \equiv {\bm x}_{\perp, 0} + 2\tau \bar {\bm p}_{\perp},\, \langle {\bm p}_{\perp}\rangle = \bar{\bm p}_{\perp}. \label{Sh.7}
\end{eqnarray}
Here, $a$ is a (real-valued) constant determining the packet's spreading with ``time'' $2\tau$.

The second example is an Airy beam \cite{Airy}, which can be obtained by combining both approaches mentioned in the beginning of this section. A Fourier expansion for the function $\Phi$ obeying Eq.(\ref{Sh.5}) is\footnote{From now on we will use for simplicity a Lorentz non-invariant integration measure $d^3 p/(2\pi)^3$, so that $|\psi ({\bm p})|^2$ is not invariant either.}
\begin{eqnarray}
& \displaystyle \Phi (\tau, {\bm x}_{\perp}) = \int \frac{d^2 {\bm p}_{\perp}}{(2\pi )^2} \Phi ({\bm p}_{\perp}) e^{-i\tau {\bm p}_{\perp}^2 + i {\bm p}_{\perp} {\bm x}_{\perp}},  \label{Sh.8}
\end{eqnarray}
and when the Fourier transform is chosen as $\Phi ({\bm p}_{\perp}) = \text{const}\, \exp\{i(p_x^3 + p_y^3)/3\}$, we arrive at the ``pure Airy'' beam:
\begin{eqnarray}
& \displaystyle \Phi (\tau, {\bm x}_{\perp}) = \text{const}\, \text{Ai}(x - \tau^2) \text{Ai}(y - \tau^2)\cr & \qquad \qquad \displaystyle \times \exp \left\{i \tau (x + y) - i \frac{4}{3} \tau^3\right\}. \label{Sh.10}
\end{eqnarray}
The square of this maximizes in a vicinity of $\text{Ai} (-1)$, which gives a parabolic motion, $x_m, y_m \approx -1 + \tau^2$. However similar to a ``pure Bessel'' state with OAM, which is (see e.g., Refs. \cite{Bliokh-11, me})
\begin{eqnarray}
\displaystyle \Phi ({\bm p}_{\perp}) \propto \delta (p_{\perp} - \kappa) e^{i \ell \phi_p}\ \Rightarrow \Phi (\tau, {\bm x}_{\perp}) \propto J_{\ell} (\kappa \rho) e^{i\ell\phi_r}, \label{Sh.11}
\end{eqnarray}
solution (\ref{Sh.10}) is non-normalizable in infinite volume since Airy functions are not square integrable. The last is clear already from the fact that the Fourier transform is non-normalizable either. That is why interpretation of this parabolic behavior is quite limited; see below and \cite{Airy}.

We would like to emphasize that these solutions to the 2D Schr\"odinger equation are exact and do not require paraxial limit. After substituting this back into Eq.(\ref{Sh.4}), one obtains exact solutions to the Klein-Gordon equation. Developing this technique, one could also construct wave-packet solutions of Bessel and Airy types, however, we prefer to obtain these ones within the second approach developed in the next section.

\section{Relativistic particles' wave-packets and Airy states}

To begin with, let us consider a case without OAM. The simplest choice in momentum representation is a widely-used Gaussian wave packet in transverse plane:
\begin{eqnarray}
&& \displaystyle \psi ({\bm p};\kappa, p_{\parallel}, \sigma) = \cr && \displaystyle = \frac{(2 \pi)^{3/2}}{\sqrt{L} \sigma}\, \delta (p_z - p_{\parallel}) \exp \left\{-\frac{1}{4\sigma^2} ({\bm p}_{\perp} - {\bm \kappa})^2\right\}, \label{Eq8.0}
\end{eqnarray}
which is normalized according to $\langle\psi|\psi \rangle = \int d^3 p |\psi ({\bm p})|^2/(2 \pi)^3 = 1$, and has a ``plane-wave limit'' in the following sense:
\begin{eqnarray}
& \displaystyle \lim_{\sigma \rightarrow 0}|\psi ({\bm p};\kappa, p_{\parallel}, \sigma)|^2 = (2\pi)^3 \delta ({\bm p} - {\bar{\bm p}}),\, {\bar{\bm p}} = \{{\bm \kappa}, p_{\parallel}\} \label{Eq8.0a}
\end{eqnarray}
Here, regularization of the delta-function squared is done in the usual way: $(\delta (p_z - p_{\parallel}))^2 \rightarrow L/2\pi\,\delta (p_z - p_{\parallel})$, with $L$ being some (large) length of the normalization cylinder. 


These states are obviously orthogonal in longitudinal momenta but have some overlap in transverse momenta:
\begin{eqnarray}
\displaystyle \int \frac{d^3 p}{(2\pi)^3} \psi^*({\bm p};{\bm \kappa}^{\prime}, p_{\parallel}^{\prime}, \sigma^{\prime}) \psi({\bm p};{\bm \kappa}, p_{\parallel}, \sigma) = \cr \displaystyle = \frac{4\pi}{L} \frac{\sigma \sigma^{\prime}}{\sigma^2 + (\sigma^{\prime})^2}\, \delta (p_{\parallel} - p_{\parallel}^{\prime}) \exp \left\{-\frac{({\bm \kappa} - {\bm \kappa}^{\prime})^2}{4 (\sigma^2 + (\sigma^{\prime})^2)}\right\}, \label{Eq8.0c}
\end{eqnarray}
Then, as long as the transverse momenta are not fixed, this state is no longer monochromatic. Indeed, if the momentum distribution is sharp, $\sigma \ll \kappa$, we can expand the energy as a function of the transverse momenta as follows:
\begin{eqnarray}
&& \displaystyle \varepsilon ({\bm p}_{\perp}) \approx \varepsilon_0 + {\bm u}_{\perp} ({\bm p} - {\bm \kappa}) + \cr && \displaystyle + \frac{1}{2\varepsilon_0}\left (\delta_{ij} - u_{\perp,i}u_{\perp,j}\right ) ({\bm p} - {\bm \kappa})_i({\bm p} - {\bm \kappa})_j,\cr && \qquad \displaystyle \varepsilon_0 : = \varepsilon ({\bm \kappa}),\ {\bm u}_{\perp} = \frac{{\bm \kappa}}{\varepsilon_0}, \label{Eq8.2}
\end{eqnarray}
which yields
\begin{eqnarray}
&& \displaystyle \langle\psi|\hat{H}|\psi\rangle = \int \frac{d^3 p}{(2\pi)^3}\,\varepsilon ({\bm p}) |\psi ({\bm p})|^2 \approx \cr && \displaystyle  \approx \varepsilon_0 + \frac{\sigma^2}{\varepsilon_0}\left (1 - \frac{1}{2} {\bm u}_{\perp}^2\right ) = \cr && \displaystyle = \varepsilon_0 \left (1 + {\bm u}_{\perp}^2 \left (\frac{\sigma}{\kappa}\right )^2 \left (1 - \frac{1}{2} {\bm u}_{\perp}^2\right ) \right ). \label{Eq8.1}
\end{eqnarray}
In all practical cases, difference of this from $\varepsilon_0$ is negligibly small (see characteristic values of $u_{\perp}$ below). 

When calculating the wave function in the coordinate representation, the integral over transverse momenta can be evaluated with the use of the following formula
\begin{eqnarray}
&& \displaystyle \int d^n x \exp \left \{- x_i a_i - \frac{1}{2} x_i B_{ij} x_j\right \} = \cr && \qquad \displaystyle = \frac{(2\pi)^{n/2}}{\sqrt{\det B}} \exp\left \{\frac{1}{2}a_i B^{-1}_{ij} a_j\right \}, \label{Eq8.2a}
\end{eqnarray}
with $B_{ij}$ being some $n\times n$ non-singular matrix. The result is found to be
\begin{widetext}
\begin{eqnarray}
& \displaystyle \psi (x) = \int \frac{d^3 p}{(2\pi)^3}\,\psi({\bm p}) e^{-ipx} = \frac{1}{\sqrt{2\pi L \det B}\ \sigma} \exp \Big\{-i\varepsilon_0 t + i z p_{\parallel} + i {\bm \kappa}{\bm r}_{\perp} - \cr & \displaystyle - \frac{1}{2} \left (\frac{\delta_{ij}}{(2\sigma^2)^{-1} + it/\varepsilon_0} + \frac{it}{\varepsilon_0} \frac{{\bm u}_{\perp,i}{\bm u}_{\perp,j}}{\det B}\right ) ({\bm r}_{\perp} - {\bm u}_{\perp} t)_i({\bm r}_{\perp} - {\bm u}_{\perp} t)_j \Big\} \label{Eq8.3}
\end{eqnarray}
\end{widetext}
i.e. represents a wave packet in the transverse plane while being delocalized (periodic) both in z and t. Here, we have used the following denotations:
\begin{eqnarray}
&& \displaystyle B_{ij} = \delta_{ij}\left (\frac{1}{2\sigma^2} + \frac{it}{\varepsilon_0}\right ) - u_{\perp,i} u_{\perp,j} \frac{it}{\varepsilon_0}, \cr && \displaystyle \det B = \left (\frac{1}{2 \sigma^2} + \frac{it}{\varepsilon_0}\right ) \left (\frac{1}{2\sigma^2} + \frac{it}{\varepsilon_0} \left (1 - {\bm u}_{\perp}^2\right )\right ). \label{Eq8.3a}
\end{eqnarray}
With an accuracy of high-order terms, this state is essentially \textit{quasi-classical}, since the wave packet center moves along a classical trajectory with the zero initial conditions:
\begin{eqnarray}
\displaystyle \langle{\bm r}_{\perp}\rangle = \int d^3 x\, {\bm r}_{\perp}\, |\psi (x)|^2 = {\bm u}_{\perp} t + O ({\bm u}_{\perp}^2),\, \langle{\bm p}_{\perp}\rangle = {\bm \kappa} \label{Eq8.3b}
\end{eqnarray}
Actually, the smallness of ${\bm u}_{\perp}^2$ as compared to unity just means that the transverse motion of the wave packet stays non-relativistic while staying relativistic along z axis. For $200-300$ KeV vortex- and Airy electrons experimentally realized so far, the values of $\kappa$ are less than $10$ KeV (see e.g., \cite{Angstrom}) which gives the following estimate:
$$
{\bm u}_{\perp}^2 < 10^{-4}.
$$

With the same accuracy we can calculate the dispersions and find that the uncertainty relations are minimized only in the initial moment of time:
\begin{eqnarray}
&& \displaystyle \langle (\Delta{\bm r}_{\perp})^2\rangle = \langle{\bm r}_{\perp}^2\rangle - \langle{\bm r}_{\perp}\rangle^2 = \frac{1}{2\sigma^2} + 2\sigma^2 \left (\frac{t}{\varepsilon_0}\right )^2, \cr && \displaystyle \langle (\Delta{\bm p}_{\perp})^2\rangle = \langle{\bm p}_{\perp}^2\rangle - \langle{\bm p}_{\perp}\rangle^2 = 2 \sigma^2,\cr && \displaystyle \langle (\Delta{\bm r}_{\perp})^2\rangle \langle (\Delta{\bm p}_{\perp})^2\rangle = 1 + \left (2\sigma^2 \frac{t}{\varepsilon_0}\right )^2, \label{Eq8.3c}
\end{eqnarray}
while $\langle (\Delta x)^2\rangle_{t = 0} \langle (\Delta p_x)^2\rangle = 1/4$. 

Such approximate (paraxial) states are widely used, for instance, in the theory of neutrino oscillations (see e.g., \cite{Akhmedov_09, Akhmedov_10}). In more recent studies, these wave-packets have been generalized in the covariant way, so that they acquire different longitudinal- and transverse dispersion rates \cite{Naumov_10, Naumov_10_2, Naumov_14}. 

In terms of null-plane variables, a similar wave-packet, which is now normalized on a hyperplane $\xi = \text{const}$\footnote{So that $d^4 x = \frac{1}{2}d\xi d\tilde{\xi}d^2{\bm r}_{\perp}, d^3 x \equiv d\tilde{\xi}d^2{\bm r}_{\perp}$}, can be obtained from Eq.(\ref{Sh.6}) of the previous section:
\begin{eqnarray}
&& \displaystyle \psi (x) = \sqrt{\frac{a}{\pi L}} \frac{1}{a + i\xi/\lambda} \exp\Big\{-\frac{i}{2}\lambda\tilde{\xi}-\frac{i}{2\lambda} m^2 \xi +  \cr && \displaystyle + i {\bm \kappa} \left ({\bm r}_{\perp} -\frac{1}{2}\bar{{\bm r}}_{\perp}\right ) - \frac{1}{2} \frac{1}{a + i\xi/\lambda}\left ({\bm r}_{\perp} - \bar{{\bm r}}_{\perp}\right )^2\Big\} \label{Eq8.4}
\end{eqnarray}
with
\begin{eqnarray}
& \displaystyle
\bar{{\bm r}}_{\perp} = {\bm r}_{\perp, 0} + {\bm \kappa} \frac{\xi}{\lambda},\ \langle{\bm r}_{\perp}\rangle = \bar{{\bm r}}_{\perp} \label{Eq8.4a}
\end{eqnarray}
representing a sort of the relativistic particle's ``classical trajectory'' in the null-plane-variables formalism. Note that now we consider a general case of non-zero initial conditions for the coordinates. 

A distinctive feature of these states (which in terms of Refs. \cite{B-G, Unbound, Unbound_2} also could be called \textit{the squeezed partially coherent states}) as compared to (\ref{Eq8.3}) is that they represent an \textit{exact solution} of the Klein-Gordon equation. Its Fourier-transform is found as:
\begin{eqnarray}
&& \displaystyle \psi (p) = \int d^4 x\, \psi (x) e^{ipx} =  \cr && \displaystyle = (2\pi)^3 2(np) \sqrt{\frac{a}{\pi L}}\, \delta (p^2 - m^2)\, \delta \left((np) - \lambda\right) \cr && \displaystyle \times \exp\left\{- i {\bm r}_{\perp, 0} \left ({\bm p}_{\perp} -\frac{1}{2}{\bm \kappa}\right ) - \frac{1}{2} a \left ({\bm p}_{\perp} - {\bm \kappa}\right )^2\right\} \equiv \cr && \qquad \qquad \displaystyle \equiv 2\pi (np) \delta \left(p^2 - m^2\right) \psi ({\bm p}) = \cr && \qquad \qquad \displaystyle = 2\pi \delta \left ((\tilde{n}p) - \frac{{\bm p}_{\perp}^2 + m^2}{(np)}\right) \psi ({\bm p}). \label{Eq8.5}
\end{eqnarray}
which is similar to (\ref{Eq8.0}). Note that the averages are calculated in momentum representation with the measure $d^3 p = \frac{1}{2}d(np) d^2 {\bm p}_{\perp}$ (i.e. on a plane $(\tilde{n}p) = \text{const}$): $
\langle{\bm p}_{\perp}\rangle = \int d^3p\, {\bm p}_{\perp} |\psi ({\bm p})|^2/(2\pi)^3 = {\bm \kappa}$. When proving this identity, it is convenient to take the normalization factor appearing when squaring the delta-function of a light-cone variable twice smaller as compared to the Cartesian analogue:
\begin{eqnarray}
&& \displaystyle \delta \left ((np) - \lambda\right ) = \int\frac{d\tilde{\xi}/2}{2\pi} \exp \left\{i\frac{\tilde{\xi}}{2} \left ((np) - \lambda\right)\right\} \cr && \qquad \qquad \displaystyle \rightarrow \int \frac{1}{2}\frac{d\tilde{\xi}}{2\pi} = \frac{1}{2} \frac{L}{2\pi}. \label{Eq8.5b}
\end{eqnarray}

Similarly to (\ref{Eq8.3c}), one can calculate the dispersions and find the uncertainty relations. We arrive at the following:
\begin{eqnarray}
&& \displaystyle \langle (\Delta{\bm r}_{\perp})^2\rangle = a + \frac{1}{a} \left (\frac{\xi}{\lambda}\right )^2,\ \langle (\Delta{\bm p}_{\perp})^2\rangle = \frac{1}{a},\cr && \displaystyle \langle (\Delta{\bm r}_{\perp})^2\rangle \langle (\Delta{\bm p}_{\perp})^2\rangle = 1 + \left (\frac{\xi}{a \lambda}\right )^2. \label{Eq8.6}
\end{eqnarray}

As is clearly seen from Eq.(\ref{Eq8.0}), the mean value of the OAM, $\langle \hat{L}_z \rangle$, equals zero for paraxial states. However, this is not the case for non-paraxial states from Eq.(\ref{Eq8.4}), and the result is:
\begin{eqnarray}
& \displaystyle \langle \hat{L}_z\rangle = [{\bm r}_{\perp, 0} \times {\bm \kappa}]_z \equiv [\langle {\bm r}_{\perp}\rangle_{\xi=0} \times \langle {\bm p}_{\perp}\rangle]_z, \label{Eq8.6aaa}
\end{eqnarray}
thanks to the non-zero initial conditions for the coordinates. 
Nevertheless this OAM has ``kinematic'' nature, as it vanishes for the zero initial conditions or the zero transverse momentum (see also \cite{Neil, Barnett_2006}). Indeed, explicit dependence on these initial conditions implies indirect dependence on time, since the moment $t = 0$ ($\xi = 0$) could be chosen absolutely arbitrarily. However for a free particle, $L_z$ is one of the exact integrals of motions, since its operator commutes with the Hamiltonian. In other words, this OAM has extrinsic nature because it appeared as a result of the choice of the quantization axis shifted from the beam's center in initial moment of time. That is why an effective (\textit{intrinsic}) value of this OAM may be said to be zero.

In a similar manner we can obtain Airy wave-packets by multiplying the Fourier transform in Eq.(\ref{Eq8.5}) on 
$$
\exp \{i b_x^3 p_x^3/3 + i b_y^3 p_y^3/3\} \equiv \exp \{(i b_x^3 p_x^3 + i b_y^3 p_y^3)/3\hbar^3\}
$$
with ${\bm b} = \{b_x, b_y\}$ being some real-valued vector characterizing initial position of the Airy beam's central peak. Note that these additional factors do not influence normalization of the states but just change their transversal overlap. Thus we still have a \textit{well normalized one-particle state}. Returning back into the configuration space, we arrive at the following:
\begin{widetext}
\begin{eqnarray}
& \displaystyle \psi (x) = 2\pi \sqrt{\frac{a}{\pi L}} \frac{1}{b_x b_y} \exp\Big\{-\frac{i}{2}\lambda\tilde{\xi} - \frac{i}{2} \xi \frac{m^2}{\lambda} + \frac{i}{2}\, {\bm \kappa}{\bm r}_{\perp, 0}  - \frac{a}{2}\kappa^2 + \frac{1}{2 b_x^3} (a + i\xi/\lambda) (x - x_0 - ia \kappa_x) + \cr & \displaystyle + \frac{1}{2 b_y^3} (a + i\xi/\lambda) (y - y_0 - ia \kappa_y) + \frac{(a + i\xi/\lambda)^3}{12} \left (\frac{1}{b_x^6} + \frac{1}{b_y^6}\right)\Big\}\times  \cr & \displaystyle \times \text{Ai}\left (b_x^{-1} \left (x -x_0 - ia\kappa_x + \frac{(a + i\xi/\lambda)^2}{4 b_x^3}\right )\right ) \text{Ai}\left (b_y^{-1}\left (y - y_0 - ia\kappa_y + \frac{(a + i\xi/\lambda)^2}{4 b_y^3}\right )\right ) \label{Eq8.6a}
\end{eqnarray}
\end{widetext}
Putting $b_x = b_y = 0$, we return to the ordinary Gaussian partially coherent states, though this is not quite obvious from the last formula.

If needed, one can also write down an \textit{approximate} (paraxial) Airy-packet solution in approach with the variables $t, z$ instead of $\xi, \tilde{\xi}$. As easy to see from the above analysis, when ${\bm u}_{\perp}^2 \ll 1$ this can be done by the following substitutions:
\begin{eqnarray}
& \displaystyle \exp\left\{-\frac{i}{2}\lambda\tilde{\xi} - \frac{i}{2} \xi \frac{m^2}{\lambda}\right\} \rightarrow \exp\left\{- i\varepsilon_0 t + i p_{\parallel} z\right\},\cr & \displaystyle a \rightarrow \frac{1}{2\sigma^2},\, \frac{\xi}{\lambda} \rightarrow \frac{t}{\varepsilon_0}. \label{Eq8.6e}
\end{eqnarray}
Moreover, this rule does not depend on the physical model of states we consider, so that these substitutions are applicable for ``switching'' between exact solutions in the null-plane variables formalism and approximate ones in the usual approach for a wide range of models. We will present another example of this rule in the next section. 

Dealing with all these wave-packet states, it is much easier to calculate averages in momentum representation in which $\hat{x} = i\partial_{p_x} + p_x \xi/(np)$, and the second term has appeared due to dispersion law, $p^2 = m^2$. The trajectory calculated with these Airy states is:
\begin{eqnarray}
& \displaystyle \langle x \rangle = \langle x \rangle_{b_x = 0} - \frac{b_x^3}{2} \left (\frac{1}{a} + 2 \kappa_x^2\right ) = X_0 + \kappa_x \frac{\xi}{\lambda},\cr & \displaystyle \langle y \rangle = \langle y \rangle_{b_y = 0} - \frac{b_y^3}{2} \left (\frac{1}{a} + 2 \kappa_y^2\right ) = Y_0 + \kappa_y \frac{\xi}{\lambda}\label{Eq8.6c}
\end{eqnarray}
with $\langle x \rangle_{b_x = 0}, \langle y \rangle_{b_y = 0}$ taken from Eq.(\ref{Eq8.4a}) and $X_0, Y_0$ are some new constants. 
Thus, despite the parabolic behavior of the Airy functions' arguments, the wave-packet's center still moves along \textit{the same quasi-classical straight path}. It is interesting to note, however, that new constants, $X_0, Y_0$, contain the Planck constant in the denominator and, therefore, have no smooth behavior in the quasi-classical limit $\hbar \rightarrow 0$. 
\begin{figure*}
\center
\includegraphics[width=14.00cm, height=5.50cm]{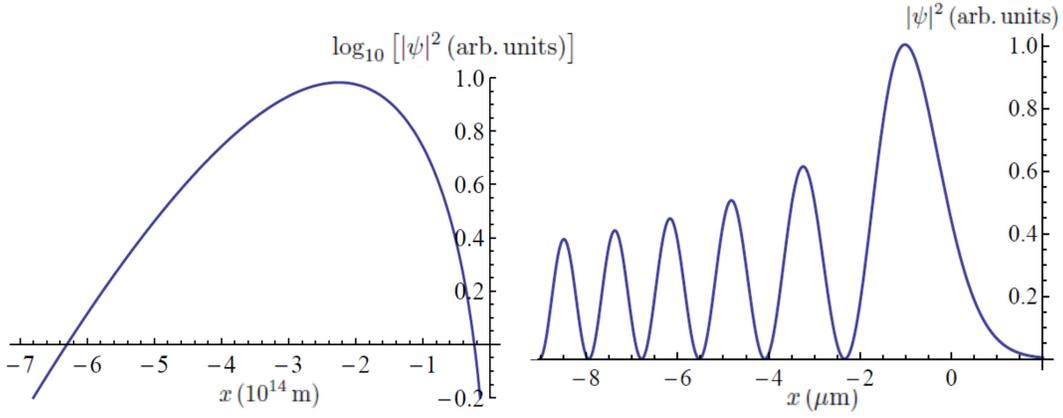}
\caption{(Color online) Probability density of the one-particle Airy wave-packet at zero distance (paraxial $t, z$ approach). Parameters: $\varepsilon_c = 200$ keV, $b_x = b_y = 1\, \mu$m , $y = \kappa_y = 0$, $\kappa_x = 0.1$ eV, $\sigma = 0.01$ eV. It has an exponentially decaying tail at macroscopic distances (left panel) while resembling a ``pure`` Airy beam at the microscopic scale (right panel). \label{Fig1}}
\end{figure*}
\begin{figure}
\center
\includegraphics[width=6.50cm, height=5.00cm]{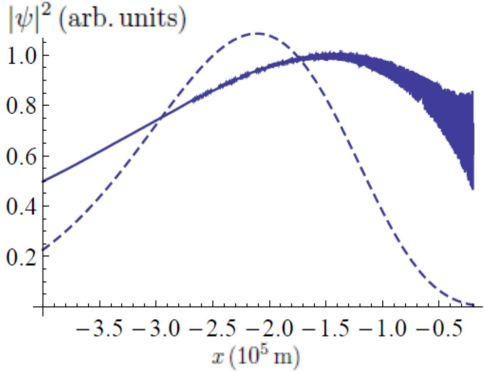}
\caption{(Color online) Airy packet's tail for different packet's widths. Solid line: $\sigma = 0.04$ eV (closer to a ``pure Airy'' beam), dashed line (decreased in $10^4$ times): $\sigma = 0.02$ eV (closer to a Gaussian beam). Here, $b_x = b_y = 1$ nm, and other parameters are the same as in Fig.\ref{Fig1}.\label{Fig2}}
\end{figure}

Furthermore, one can also calculate the dispersions, $\langle (\Delta{\bm r}_{\perp})^2\rangle$, $\langle (\Delta{\bm p}_{\perp})^2\rangle$, and make sure that they \textit{coincide} with the ones in Eq.(\ref{Eq8.6}), i.e. the Airy packet spreads with time and $z$ \textit{exactly as a Gaussian beam does}, in accordance with the uncertainty relations. In other words, the Airy wave-packet (\ref{Eq8.6a}) also represents \textit{a quasi-classical state} in a sense that it minimizes the uncertainty relations in initial moment of time (or when $\xi = 0$), despite the terms of higher powers of $\hbar$. This is also a curious example of a quasi-classical state that is non-Gaussian in the configuration space. 

We would like to emphasize that there is no contradiction here with the experimentally verified fact that the Airy packets are (almost) non-spreading revealing some diffraction-free and self-healing properties (see e.g., \cite{Airy_beam, Airy_Exp, Airy_El_Exp}). The point is that the Airy wave-packet spreads mostly \textit{at the expense of its tail}, while preserving the shape and width of the central peak (similar to the optical Airy beam: see e.g., \cite{Airy_Exp}). Indeed, an effective time when the coordinate uncertainty doubles is (compare with \cite{Gold}; we will use a ``usual'' $t, z$ approach here for simplicity):
\begin{eqnarray}
& \displaystyle T = \frac{\varepsilon_0}{2 \sigma^2} \label{Eq8.6ca}
\end{eqnarray}
For $200$ KeV electrons of the experiment \cite{Airy_El_Exp} with the coordinate uncertainty of the order of transverse size of the beam, $\sigma_x \sim 10\, \mu\text{m}$, and the corresponding momentum uncertainty $\sigma \sim 1/\sigma_x \sim 2\times 10^{-2}$ eV, we obtain the following estimate for the minimum distance needed for a packet to spread twice:
\begin{eqnarray}
& \displaystyle L = uT \sim 125\, \text{m}, \label{Eq8.6cb}
\end{eqnarray}
which exactly is of the order of the maximum distance at which measurements were carried out in Ref.\cite{Airy_El_Exp}. However registered FWHM of Airy beam at the distance $100$ m was slightly less than 10\% higher than the one right after the experimental setup, which just means that the estimate (\ref{Eq8.6cb}), being applicable for the beam as a whole, is not so for the central Airy peak. 

In Fig.\ref{Fig1} we depicted the central peak of the Airy wave-packet (right panel) and the packet's tail (left panel). For simplicity, we use the ``ordinary'' $t, z$ approach. As we consider one particle in the Universe, the tail's peak may be far away from the Airy's central peak. As can be seen, the tail peak's position is scaled with the third power of $b_x, b_y$, while the main Airy peak is scaled with the first power. So that for initial conditions, for instance, of $b_y = b_x = 1$ nm (instead of $1\, \mu\text{m}$) and the other parameters staying the same, we would obtain the same figures as in Fig.\ref{Fig1} when changing $10^{14} \rightarrow 10^{5}$ in the left panel's caption and $\mu \text{m} \rightarrow \text{nm}$ in the right panel's caption. We illustrate this in Fig.\ref{Fig2} while also showing that the limiting case of a tightly focused in configuration space state (larger values of $\sigma$) corresponds to a ``pure'' Airy beam. Note that in the non-paraxial case with $\sigma \sim \kappa$ the ``usual'' $t, z$ approach becomes inapplicable, whereas the one with the null-plane variables still works.

Now let us address the question of whether an Airy beam may possess some orbital angular momentum with respect to the z-axis \cite{OAM-Airy}. Calculating mean value of the OAM's z-projection with the Airy states we obtain the following quasi-classical result:
\begin{eqnarray}
& \displaystyle \langle \hat{L}_z \rangle \equiv \langle [\hat{{\bm r}}_{\perp} \times \hat{{\bm p}}_{\perp}]_z \rangle = \kappa_y \left (x_0 - \frac{b_x^3}{2} \left (\frac{1}{a} + 2\kappa_x^2\right )\right ) - \cr & \displaystyle - \kappa_x \left (y_0 - \frac{b_y^3}{2} \left (\frac{1}{a} + 2\kappa_y^2\right )\right ) = \cr & \displaystyle = \kappa_y X_0 - \kappa_x Y_0 \equiv \left [\langle{\bm r}_{\perp}\rangle_{\xi = 0} \times \langle{\bm p}_{\perp}\rangle\right ]_z \label{Eq8.6d}
\end{eqnarray}
We see that the OAM generally does not vanish, and it can be turned into zero only by a very special choice of the initial conditions: 
\begin{eqnarray}
& \displaystyle X_0 = Y_0 = 0 \Leftrightarrow \cr & \displaystyle x_0 = \frac{b_x^3}{2} \left (\frac{1}{a} + 2\kappa_x^2\right ),\, y_0 = \frac{b_y^3}{2} \left (\frac{1}{a} + 2\kappa_y^2\right ). \label{Eq8.6da}
\end{eqnarray}
On the other hand, this result coincides, up to notations, with Eq.(\ref{Eq8.6aaa}) calculated for an ordinary Gaussian beam. This means that such an OAM also may be treated as ``kinematic'' or extrinsic as it can be turned into zero by the choice of the quantization axis or the initial conditions. In other words, its effective value is zero as was indicated in Ref.\cite{OAM-Airy}.

However, as we will demonstrate below, vanishing mean value of the OAM itself \textit{is not sufficient} for a general statement that this wave-packet does not carry any OAM (analogously to the optical case: see Ref.\cite{Barnett_2006}). If the second moment of the OAM does not vanish, such wave-packet has some distribution in the orbital momenta space or, in other words, has finite OAM bandwidth (see e.g., \cite{Torres_2003, F-A_EPJD_2012, F-A_2012}) related to the overall number of the OAM modes carried by such a wave-packet.

\section{Relativistic particles' wave-packets with OAM}

Now let us add OAM quantized along the z-axis. The Gaussian wave-packets we considered had plane-wave limit when $\sigma \rightarrow 0$, so that OAM effectively vanishes in this case (see below)
. One can also construct normalized one-particle Gaussian wave-packets with an azimuthal-independent probability density pattern and a ``pure-Bessel'' limit:
\begin{eqnarray}
& \displaystyle \psi ({\bm p}; \kappa, p_{\parallel}, \sigma, \ell) = (2 \pi)^{3/2} (-i)^{\ell} \frac{\sqrt{2}}{\sqrt{1 + \text{erf} \left (\frac{\kappa}{\sqrt{2}\sigma}\right )}}\times \cr & \displaystyle \frac{1}{\sqrt{L} (\pi 2 \sigma^2)^{1/4}} \frac{\delta (p_z - p_{\parallel})}{\sqrt{p_{\perp}}} \exp \left \{-\frac{(p_{\perp} - \kappa)^2}{4\sigma^2} + i \ell \phi \right\}, \cr & \displaystyle
\lim_{\sigma\rightarrow 0}|\psi ({\bm p}; \kappa, p_{\parallel}, \sigma, \ell)|^2 = (2\pi)^2 \delta (p_z - p_{\parallel}) \frac{\delta (p_{\perp} - \kappa)}{p_{\perp}}\label{Eq8.7}
\end{eqnarray}
so that the momentum's azimuthal angle has no ``mean value''. We will call such wave packets Gaussian-Bessel ones. Note that they are orthogonal in OAM, $\int d^3 p\,\psi^* ({\bm p}; \ell^{\prime}) \psi ({\bm p}; \ell)/(2\pi)^3 \propto \delta_{\ell, \ell^{\prime}}$.

The coordinate representation for the wave function,
\begin{eqnarray}
& \displaystyle \psi (x) = \int \frac{d^3 p}{(2\pi)^3} \psi ({\bm p}) e^{-ipx} = \cr & \displaystyle = \frac{e^{ip_{\parallel} z + i \ell \phi_r}}{\sqrt{\pi L} (\pi 2 \sigma^2)^{1/4}\sqrt{1 + \text{erf} \left (\frac{\kappa}{\sqrt{2}\sigma}\right )}} \cr & \displaystyle \times \int \limits_0^{\infty} d p_{\perp} \sqrt{p_{\perp}}\,J_{\ell} (p_{\perp} \rho) \exp\Big\{-it\varepsilon (p_{\perp}) - \cr & \qquad \qquad \displaystyle - \frac{1}{4\sigma^2} (p_{\perp} - \kappa)^2\Big\},\label{Eq8.8}
\end{eqnarray}
represents an eigenfunction for $\hat{L}_z$ operator, and the probability density profile has a typical ``doughnut'' spatial structure with the central minimum (see e.g., \cite{Bliokh-11} and also the recent experiment \cite{Grillo}). These wave-packets represent direct generalization of the ``pure Bessel'' states, and they can be useful for analyzing problems of scattering with the twisted particles \cite{I-S}.


Even more useful objects are obtained by embedding OAM into the ``usual'' 2D Gaussian wave-packets with the plane-wave limit. This is done by multiplying the wave functions in momentum representation on the following factor: $\psi ({\bm p};\kappa, p_{\parallel}, \sigma, \ell) = (-i)^{\ell} e^{i\ell\phi_p} \psi ({\bm p};\kappa, p_{\parallel}, \sigma)$. In the null-plane variables, this will be [here, $\psi ({\bm p})$ is taken from (\ref{Eq8.5})]: 
\begin{eqnarray}
& \displaystyle \psi (x) = \int \frac{d^3 p}{(2\pi)^3} \psi ({\bm p})(-i)^{\ell} e^{i\ell\phi_p} e^{-ipx} = \cr & \displaystyle = \sqrt{\frac{a}{\pi L}} \exp \left\{-\frac{i}{2}\lambda\tilde{\xi} - i \xi \frac{m^2}{2 \lambda} -\frac{a}{2}\kappa^2 + \frac{i}{2} {\bm r}_{\perp,0} {\bm \kappa} + i\ell\varphi\right\} \cr & \displaystyle \times\int \limits_0^{\infty} dp_{\perp}p_{\perp} J_{\ell} (p_{\perp} R) \exp\left\{-p_{\perp}^2 \frac{1}{2}\left (a + i\frac{\xi}{\lambda}\right )\right\}.\label{Eq8.9}
\end{eqnarray}
Here, ${\bm R} = {\bm r}_{\perp} - {\bm r}_{\perp,0} - i a {\bm \kappa} = R \{\cos \varphi, \sin \varphi\}$, so that $\varphi$ acquires an imaginary part. The last integral is done with the help of Eq.(6.631) in \cite{Ryzh} and the final result is
\begin{widetext}
\begin{eqnarray}
& \displaystyle \psi (x) = \sqrt{\frac{a}{2 L}} \frac{R}{2} \left (a + i\frac{\xi}{\lambda}\right )^{-3/2}\exp\left\{-\frac{i}{2}\lambda\tilde{\xi} - i \xi \frac{m^2}{2 \lambda} -\frac{a}{2}\kappa^2 + \frac{i}{2} {\bm r}_{\perp,0} {\bm \kappa} -\frac{1}{4} \frac{{\bm R}^2}{a + i\xi/\lambda} + i\ell\varphi\right\} \cr & \displaystyle \times \left (I_{\frac{1}{2} (\ell - 1)} \left (\frac{1}{4} \frac{{\bm R}^2}{a + i \xi/\lambda} \right ) - I_{\frac{1}{2} (\ell + 1)} \left (\frac{1}{4} \frac{{\bm R}^2}{a + i \xi/\lambda} \right )\right ),\label{Eq8.10}
\end{eqnarray}
\end{widetext}
where $I$ denotes a modified Bessel function. This expression, which can be called a Gaussian one-particle state with the OAM (or a squeezed partially coherent state), represents \textit{an exact solution} for the Klein-Gordon equation, and in the limiting case $\ell = 0$ it coincides with the ``ordinary'' coherent states (\ref{Eq8.4}). 
 As easy to show, an approximate solution with the OAM in the parametrization (\ref{Eq8.0}) can be obtained by the same substitution (\ref{Eq8.6e}). 

It might seem that this wave function is not an eigenfunction of $\hat{L}_z$ in a general case. In particular, in the plane-wave limit when $a \rightarrow \infty$ (or $\sigma \rightarrow 0$) we have $\varphi \rightarrow \phi_{\kappa},\ {\bm R} \rightarrow - i a {\bm \kappa}$, so that the wave function does not depend on $\phi_r$, and this yields $\hat{L}_z \psi_{a \rightarrow \infty}(x) = 0$. In the opposite case of an ultra-tightly focused (in configuration space) wave packet, $a \rightarrow 0$ ($\sigma \rightarrow \infty$), we have $\varphi \rightarrow \phi_r,\ {\bm R} \rightarrow {\bm r}_{\perp}$, so that a phase vortex appears:
$$
\psi (\rho, \phi_r, z) = \psi (\rho, z) e^{i\ell \phi_r} \Rightarrow \hat{L}_z \psi_{a \rightarrow 0}(x) = \ell \psi_{a \rightarrow 0}(x).
$$
The curious fact, however, is that even in general case of finite $a$ (or $\sigma$), the mean z-component of the OAM still equals $\ell$ for both approaches (null-plane and the usual one). This could be easier seen in momentum representation where $\hat{L}_z = -i \partial_{\phi_p}$:
\begin{eqnarray}
& \displaystyle \langle\hat{L}_z\rangle = \cr & \displaystyle \int \frac{d^3 p}{(2\pi)^3} \psi^*({\bm p};\kappa, p_{\parallel}, \sigma, \ell) \hat{L}_z \psi ({\bm p};\kappa, p_{\parallel}, \sigma, \ell) = \ell,\label{Eq8.12}
\end{eqnarray}
where an extra addend, which appears due to exponent in (\ref{Eq8.0}), vanishes after integration over $\phi_p$. 

Note also that whereas the Gaussian-Bessel packets are obviously orthogonal in $\ell$ (but not in $\kappa$!), this is not the case for the 2D Gaussian wave-packets. Indeed, calculating their overlap in momentum representation we arrive at the following:
\begin{eqnarray}
& \displaystyle \int \frac{d^3 p}{(2\pi)^3}\ \psi^* ({\bm p};{\bm \kappa}^{\prime}, p_{\parallel}^{\prime},  \sigma^{\prime}, \ell^{\prime}) \psi ({\bm p}; {\bm \kappa}, p_{\parallel}, \sigma, \ell) = \cr & \displaystyle = \frac{2\pi^{3/2}}{L} (-i)^{\ell - \ell^{\prime}} e^{i(\ell - \ell^{\prime}) \tilde{\phi}} \delta (p_{\parallel} - p_{\parallel}^{\prime}) \frac{\tilde{\kappa} \sigma^2 (\sigma^{\prime})^2}{(\sigma^2 + (\sigma^{\prime})^2)^{3/2}} \cr & \displaystyle \times \exp \left \{-\left (\frac{\kappa}{2\sigma}\right)^2 - \left (\frac{\kappa^{\prime}}{2\sigma^{\prime}}\right)^2 + \frac{(\tilde{\kappa} \sigma \sigma^{\prime})^2}{2 (\sigma^2 + (\sigma^{\prime})^2)}\right\} \cr & \displaystyle \times \Big (I_{\frac{1}{2} (\ell - \ell^{\prime} - 1)} \left (\frac{(\tilde{\kappa} \sigma \sigma^{\prime})^2}{2 (\sigma^2 + (\sigma^{\prime})^2)}\right ) + \cr & \qquad \qquad \qquad \displaystyle + I_{\frac{1}{2} (\ell - \ell^{\prime} + 1)} \left (\frac{(\tilde{\kappa} \sigma \sigma^{\prime})^2}{2 (\sigma^2 + (\sigma^{\prime})^2)}\right )\Big ),
\label{Eq8.12c}
\end{eqnarray}
where the following notation is used: $\tilde{{\bm\kappa}} \equiv \tilde{\kappa} \{\cos\tilde{\phi}, \sin\tilde{\phi}\} : = {\bm \kappa}/2\sigma^2 + {\bm \kappa}^{\prime}/2(\sigma^{\prime})^2$. In the coordinate representation, non-orthogonality in OAM is clearly seen from the fact that despite the equality $d^2{\bm \rho} = d^2 {\bm R}$, both $R$ and $\varphi$ depend on $\kappa$ and $\sigma$. Note that orthogonality is recovered in the case of the zero transverse momenta: $\kappa, \kappa^{\prime} \rightarrow 0$.

We would like to stress that both, the Gaussian-Bessel wave-packets and the 2D Gaussian ones, represent quantum states with a \textit{well-defined z-component of the OAM}, dependence of $|\psi|^2$ on the azimuthal angle in the latter case notwithstanding. In fact, such a dependence has a useful interpretation in view of the uncertainty relations for the angular momentum (see e.g, \cite{OAM_unc}):
\begin{eqnarray}
& \displaystyle \langle(\Delta L_z)^2\rangle \langle (\Delta \sin \phi_r)^2 \rangle \geq \frac{1}{4} \langle \cos\phi_r \rangle^2.\label{Eq8.13}
\end{eqnarray}
Indeed, it is a lack of this dependence for Gaussian-Bessel wave packets as well as for the ``pure-Bessel'' states that leads to their zero uncertainties of the OAM, $\langle(\Delta L_z)^2\rangle = \langle \hat{L}_z^2 \rangle - \langle \hat{L}_z \rangle^2 = 0$, and finite uncertainties of the azimuths, $\langle (\Delta \sin \phi_r)^2 \rangle = \langle(\Delta \cos \phi_r)^2\rangle = \langle \sin^2\phi_r \rangle = \langle \cos^2\phi_r \rangle = 1/2$. At the same time, for 2D Gaussian wave-packets we have a \textit{finite uncertainty} of the OAM -- a fact, which is closely related to their non-orthogonality. And as can be calculated using, for instance, the paraxial states, the OAM-dispersion is (note that here ${\bm r}_{\perp, 0} = 0$)
\begin{eqnarray}
& \displaystyle
\langle(\Delta L_z)^2\rangle = \left (\frac{\kappa}{2 \sigma}\right )^2, \label{Eq8.17}
\end{eqnarray}
while uncertainties for the azimuths stay the same. The r.h.s. of (\ref{Eq8.13}) in these cases is just $0$. Note that as the 2D Gaussian wave-packets represent $2\pi$-periodic functions of $\phi$, the $\hat{L}_z$-operator is still self-adjoint on the space of these functions. 

There exist alternative representations for the uncertainty relations (\ref{Eq8.13}), where the l.h.s. contains dispersion of the azimuthal angle itself, while the r.h.s. gets an additional addend (see e.g., \cite{Barnett_1990}). Note that these uncertainty relations were successfully checked experimentally in optics (see e.g., \cite{F-A_2004, F-A_2011}).

As the ``almost plane-wave'' states correspond to narrow packets in the momentum space, $\sigma \ll \kappa$, the OAM-dispersion gets bigger in this case, $\langle(\Delta L_z)^2\rangle \gg 1$. Consequently, effective OAM may be said to vanish in the plane-wave limit, which is a very natural result. 
On the other hand, the formula (\ref{Eq8.17}) might seem to reveal such a property of the tightly-focused scalar matter waves as to acquire \textit{induced OAM}. This feature is known to exist for vectorial laser beams appearing thanks to the spin-orbital coupling (see e.g., \cite{F_2007}), however, in the scalar case under consideration it is not the OAM's mean value which increases when the beam gets focused, but \textit{the OAM bandwidth} (see e.g., \cite{Torres_2003}) just gets narrower so that the OAM-uncertainty decreases.
\begin{figure}
\center
\includegraphics[width=8.70cm, height=4.40cm]{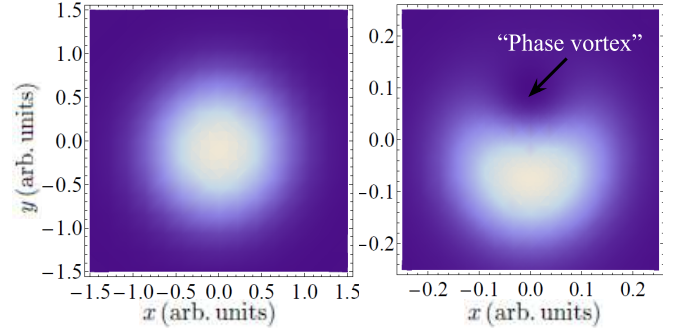}
\caption{(Color online) Probability density $|\psi|^2$ (in arbitrary units) for a 2D Gaussian beam with OAM at zero time instant (paraxial $t, z$ approach) for $\ell = 1, \phi_{\kappa} = 0$ ($\kappa_y = 0$). Left panel: $\sigma/\kappa = 0.1$ (Gaussian limit), right panel: $\sigma/\kappa = 0.7$. An azimuthally symmetric distribution with a phase vortex is recovered in the formal limit $\sigma \gg \kappa$.\label{Fig3}}
\end{figure}
\begin{figure*}
\center
\includegraphics[width=18.20cm, height=4.50cm]{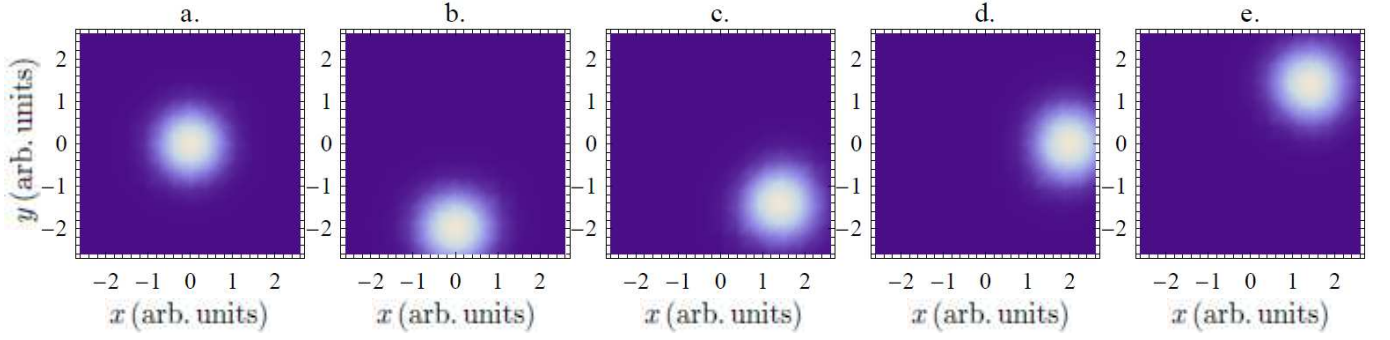}
\caption{(Color online) Probability density $|\psi|^2$ (in arbitrary units) for a 2D Gaussian beam with OAM at zero time instant (paraxial $t, z$ approach) and $\sigma/\kappa = 0.1$. a.: $\ell = 0, \phi_{\kappa} = 0$ ($\kappa_y = 0$), b.: $\ell = 20, \phi_{\kappa} = 0$ ($\kappa_y = 0$), c.: $\ell = 20, \phi_{\kappa} = \pi/4$ \textbf{($\kappa_x = \kappa_y$)}, d.: $\ell = 20, \phi_{\kappa} = \pi/2$ ($\kappa_x = 0$), e.: $\ell = 20, \phi_{\kappa} = 3 \pi/4$ ($\kappa_x = - \kappa_y$). The beam's center moves to the direction of a vector ${\bm \kappa} \times {\bm e}_z$. \label{Fig4}}
\end{figure*}

The Gaussian states with OAM, Eq.(\ref{Eq8.10}), also illustrate that the Berry's statement that there is no direct connection between phase vortices (points where $|\psi|^2$ vanishes and the phase stays undetermined) and the presence of some OAM \cite{Berry} \textit{stays valid for massive particles} as well. While a typical ``pure-Bessel'' state or a Gaussian-Bessel packet has a ``doughnut'' spatial profile, the states being considered here remain practically Gaussian when $\sigma \ll \kappa$ or even when $\sigma \lesssim \kappa$ (see Fig.\ref{Fig3}), as the ``pure Bessel'' case corresponds to the formal limit $\sigma \gg \kappa$, where the Gaussian in momentum space turns into the delta-function. In the latter case, however, paraxial states are no longer applicable, and the exact solutions (\ref{Eq8.10}) should be used instead. As the OAM increases, the beam's center moves to the direction of a vector ${\bm \kappa} \times {\bm e}_z$, as Fig.\ref{Fig4} shows. 
Note that the density profile in Fig.\ref{Fig4} stays azimuthally-symmetric (around the central peak) for any ratio between $\kappa_x$ and $\kappa_y$, which is different from the Hermite-Gaussian modes and similar to the Laguerre-Gaussian ones.

Let us now note that though a Gaussian wave-packet with the OAM and the Airy states are quite different in the configuration space, in the momentum space their wave functions differ from each other only in the general complex phase, which does not change the state's norm. Mathematically speaking, these states belong to the same class of functions having quasi-classical averages. That is why, \textit{an Airy wave-packet also has finite quantum uncertainty of the OAM} or the finite OAM bandwidth (in terms of Refs.\cite{Torres_2003, F-A_EPJD_2012, F-A_2012}). Direct calculations with the paraxial Airy wave-packet in the general case of the non-zero initial conditions yield the following result:
\begin{widetext}
\begin{eqnarray}
& \displaystyle
\langle(\Delta L_z)^2\rangle|_{Airy} = \left (\frac{\kappa}{2 \sigma}\right )^2 + \sigma^2 {\bm r}_{\perp, 0}^2 + [{\bm r}_{\perp, 0} \times {\bm \kappa}]_z^2 + 2 b_x^3 \left (y_0 \kappa_x \kappa_y (3\sigma^2 + \kappa_x^2) - x_0 (\sigma^2 + \kappa_x^2) (\sigma^2 + \kappa_y^2)\right ) + \cr & \displaystyle + 2 b_y^3 \left (x_0 \kappa_x \kappa_y (3\sigma^2 + \kappa_y^2) - y_0 (\sigma^2 + \kappa_x^2) (\sigma^2 + \kappa_y^2)\right ) - 2 b_x^3 b_y^3 \kappa_x \kappa_y (3\sigma^2 + \kappa_x^2) (3\sigma^2 + \kappa_y^2) + \cr & \displaystyle + b_x^6 (\sigma^2 + \kappa_y^2) (3\sigma^4 + 6\sigma^2\kappa_x^2 + \kappa_x^4) + b_y^6 (\sigma^2 + \kappa_x^2) (3\sigma^4 + 6\sigma^2\kappa_y^2 + \kappa_y^4). \label{Eq8.18aaad}
\end{eqnarray}
\end{widetext}
As usual, in order to obtain a similar expression for ``exact'' states it is enough to substitute $\sigma^2 \rightarrow 1/2a$. As we discussed earlier, choice of the zero initial conditions, ${\bm r}_{\perp, 0} = 0$, turns OAM of the ordinary Gaussian beam into zero, and the experimentally observable OAM bandwidth is just $\kappa/2\sigma$. Let us check now whether this trick works for Airy states as well. According to Eq.(\ref{Eq8.6da}), the special choice of the initial conditions, $x_0 = b_x^3 (\sigma^2 + \kappa_x^2), y_0 = b_y^3 (\sigma^2 + \kappa_y^2)$, turns OAM of the Airy beam into zero. The OAM uncertainty from Eq.(\ref{Eq8.18aaad}) in this case becomes:
\begin{eqnarray}
& \displaystyle
\langle(\Delta L_z)^2\rangle|_{Airy} = \left (\frac{\kappa}{2 \sigma}\right )^2 - 8 b_x^3 b_y^3 \sigma^4 \kappa_x \kappa_y + \cr & \displaystyle + 2 b_x^6 \sigma^2 (\sigma^2 + 2\kappa_x^2) (\sigma^2 + \kappa_y^2) + \cr & \displaystyle + 2 b_y^6 \sigma^2 (\sigma^2 + 2\kappa_y^2) (\sigma^2 + \kappa_x^2). \label{Eq8.18aaada}
\end{eqnarray}
Thus, the vector ${\bm b}$ gives non-vanishing ``dynamical'' contribution to the observable OAM bandwidth. Qualitatively, this result could have been expected as a transverse profile of the Airy beam is highly azimuthally-asymmetric (see figures in e.g., Ref.\cite{Airy_El_Exp}), unlike the simple Gaussian packet. Consequently, its azimuthal dispersion, $\langle(\Delta \phi_r )^2\rangle$, does not coincide with that of the Gaussian packet either. As this dispersion is connected with the one of the OAM by the uncertainty relations, the value of $\langle(\Delta L_z)^2\rangle$ also should differ from its value for ${\bm b} = 0$, (\ref{Eq8.17}).

On the other hand, the finite OAM uncertainty of the Gaussian beam (\ref{Eq8.17}) may be explained in the way that the quantization (z) axis does not coincide with the mean trajectory $\langle {\bm r}\rangle$. If this were the case, the transverse momentum $\kappa$ would turn into zero along with the OAM uncertainty\footnote{Note that while for paraxial states one cannot put $\kappa = 0$, 
this could be easily done for exact states with the light-cone variables.}. As we mentioned above, the Gaussian states become orthogonal in OAM in this case. That is why in terms of Ref.\cite{Neil} such an OAM-bandwidth could be called \textit{extrinsic}. In optics, such (almost) Gaussian states of photons with the OAM and finite OAM-uncertainty are obtained experimentally by using diffraction gratings with a fork dislocation shifted from the beam's center (see e.g., \cite{Mair_2001}; compare Fig.4 there with Fig.\ref{Fig3} here), so that the transverse momentum becomes finite. These states turned out to be very useful for purposes of the quantum entanglement in OAM of photons produced in the parametric down-conversion process (see e.g., \cite{Mair_2001, F-A_EPJD_2012, F-A_2012}). Roughly speaking, it is a finite OAM-uncertainty that allows one to successfully manipulate the OAM-entangled photons.

As is seen from Eq.(\ref{Eq8.18aaada}), the OAM dispersion could easily be much larger than unity, so that an Airy state, in fact, \textit{carries some OAM modes} with their overall number of the order of $\langle(\Delta L_z)^2\rangle$. This finiteness of the OAM bandwidth makes Airy electrons and other Airy particles potentially useful for experiments with the OAM-entangled particles (see optical examples e.g., in Refs.\cite{Mair_2001, F-A_2012}) or with the particles correlated in their OAM- vs. azimuthal-angle-distributions (see optical example e.g., in Ref.\cite{F-A_2011}). Another remarkable feature of the Airy wave-packet is that, unlike in the Gaussian case, its OAM uncertainty does not turn into zero simultaneously with the transverse momentum. As is seen, the r.h.s neither of Eq.(\ref{Eq8.18aaad}) nor of Eq.(\ref{Eq8.18aaada}) vanishes when $\kappa = 0$, so that the Airy beam's OAM-bandwidth may be said to have \textit{intrinsic} contributions. In the case of the zero OAM, $\langle \hat{L}_z\rangle = \kappa = 0$, this is just 
\begin{eqnarray}
& \displaystyle
\langle(\Delta L_z)^2\rangle|_{Airy} = 2\sigma^6 (b_x^6 + b_y^6), \label{Eq8.18aaadab}
\end{eqnarray}
and, unlike the extrinsic part (\ref{Eq8.17}), it increases for tightly-focused (in configuration space) beams. On the other hand, dependence of (\ref{Eq8.18aaad}) upon a choice of the quantization axis also means that the OAM distribution is spatially inhomogeneous (see also \cite{OAM-Airy}). That is why the Airy beams, similarly to the Bessel beams or the Laguerre-Gaussian ones \cite{Mono}, \textit{also can be used for trapping and rotating micro-particles} if the sizes of these particles are smaller than the beam's effective width, or even for rotation of the much bigger objects -- similar to the experiment of Ref.\cite{Liu}.

Finally, recalling similarity of the Airy states and the Gaussian ones, one can consider a (not necessarily Gaussian) wave-packet in momentum representation with some arbitrary complex phase:
\begin{eqnarray}
& \displaystyle
\psi ({\bm p}) \rightarrow \psi ({\bm p}) \exp \left\{i g ({\bm p})\right\}, \label{Eq8.18}
\end{eqnarray}
with $\psi ({\bm p})$ being the OAM-less wave-function [say, Eq.(\ref{Eq8.5})]. Then if the phase $g ({\bm p})$ represents a sum of the OAM-term, $\ell \phi_p$, and another function, $f ({\bm p})$, the OAM of the resultant state will, in turn, represent a sum:
\begin{eqnarray}
& \displaystyle
\langle \hat{L}_z \rangle = \ell + \langle f^{(1)}_{\phi_p} ({\bm p})\rangle,\label{Eq8.19}
\end{eqnarray}
where $f^{(1)}_{\phi}$ means derivative over $\phi_p$. For example by choosing $f ({\bm p})$ according to the Airy case and taking the same Gaussian envelope, one can obtain an Airy-Bessel wave-packet with the OAM, which would be a massive generalization (exact solution in the light-cone variables) of the corresponding paraxial beam in optics \cite{Airy_beam, AB_2010}. It is easy to check that in this case the mean value $\langle f^{(1)}_{\phi_p} ({\bm p})\rangle$ matches Eq.(\ref{Eq8.6d}), so that OAM of such a beam, being formally additive, has an effective value of just $\ell$, and its OAM bandwidth is also finite. From experimental point of view, these Airy-Bessel wave-packets could be realized by transmitting an (electron) beam through two spatially separated gratings (holograms), the first of which, having a fork dislocation, would induce OAM and the second one would induce the cubic phase (or vice versa). It is also interesting to note that a Bessel beam or a Gaussian beam with the OAM must not change its mean OAM value (while changing its spatial structure and the OAM bandwidth) when transmitting through an Airy grating. 

Drawing an analogy to the quantum optics, such Airy-Bessel electrons could be useful for experiments dealing with entanglement in the OAM. While a Bessel electron has zero OAM-uncertainty, which makes it useless for entanglement purposes, its experimentally realized OAM values can be as high as $\sim 100 \hbar$ \cite{McMorran, Uchida_2012}. For such ``highly twisted'' particles, it is an Airy grating that may be used to enrich the OAM spectrum, making it broader while effectively preserving the average OAM value. As can be shown, in a scattering process (say, in the Compton scattering) the finiteness of the OAM bandwidth is crucially important for OAM-entanglement of the final particles to occur, exactly as in the parametric down-conversion process.

\section{Generalization for a particle in plane electromagnetic wave}

Solutions we have discussed can be generalized to a case when there is some background electromagnetic field whose potential depends on the null-plane variables only. The simplest case here is a plane electromagnetic wave running in the negative z-direction:
\begin{eqnarray}
& \displaystyle A^{\mu} \equiv A^{\mu}(\xi),\ \partial_{\mu}A^{\mu} = ({\bm n} {\bm A}) = 0,\, {\bm n} = \{0, 0, -1\} \cr & \displaystyle \Rightarrow A = \{0, A_x (\xi), A_y (\xi), 0\},  \label{Eq9.0}
\end{eqnarray}
Usually, for such a geometry the well-known Volkov states are used (see e.g., \cite{BLP, B-G, Ritus, DiPiazza}), which closely resemble the plane-wave ones since they are characterized with the four-quasi-momentum going to the ``bare'' four-momentum when the field is switched off. At the same time, it was indicated some time ago that there are some non-Volkov states with a different set of quantum numbers \cite{Non-Volkov} and, in particular, there are some quantum states of an electron with the OAM for such a configuration \cite{me}. 

Here, we develop the idea that owing to the fact that potential $A$ depends on the light-cone variables only, the 4D Klein-Gordon equation can be mapped to the 2D Schr\"odinger equation, as discussed by Bagrov and Gitman in Ref.\cite{Non-Volkov}. Taking the appropriate solutions for the latter, which is much easier to do, one can obtain corresponding exact solutions for the former. The 2D Schr\"odinger equation, as we mentioned in Sec.2, has solutions of the wave-packet types, in particular, of the Airy beam and some others.  

The Klein-Gordon equation in the null-plane variables formalism,
\begin{eqnarray}
& \displaystyle \left ((\hat{p}^{\mu} - e A^{\mu})^2 - m^2 \right )\Psi (\xi, \tilde{\xi}, {\bm r}_{\perp}) = 0,  \label{Eq9.1}
\end{eqnarray}
after the substitution
\begin{eqnarray}
& \displaystyle \Psi (\xi, \tilde{\xi}, {\bm r}_{\perp}) = \exp \left \{-\frac{i}{2}\lambda \tilde{\xi} - \frac{i}{2\lambda} \int d\xi\, (m^2 + e^2 {\bm A}^2)\right\}\cr & \displaystyle \times \psi (\xi, {\bm r}_{\perp}),  \label{Eq9.2}
\end{eqnarray}
gets the following form
\begin{eqnarray}
& \displaystyle \left (2i\lambda \partial_{\xi} - \hat{\bm p}^2_{\perp} + 2e ({\bm A}\hat{\bm p}_{\perp}) \right ) \psi (\xi, {\bm r}_{\perp}) = 0,  \label{Eq9.2a}
\end{eqnarray}
Then we observe that in terms of the new radius-vector (here we follow mostly Ref.\cite{Non-Volkov}; see also \cite{me}),
\begin{eqnarray}
& \displaystyle {\bm {\mathfrak R}} = {\bm r}_{\perp} + \frac{e}{\lambda} \int d\xi\, {\bm A},  \label{Eq9.3}
\end{eqnarray}
the last equation is just a free 2D Schr\"odinger equation:
\begin{eqnarray}
& \displaystyle \psi (\xi, {\bm r}_{\perp}) \equiv \psi \left (\xi, {\bm {\mathfrak R}} = {\bm r}_{\perp} + \frac{e}{\lambda} \int d\xi\, {\bm A}\right ) \Rightarrow \cr & \displaystyle \Rightarrow  \left (2i\lambda \partial_{\xi} - \hat{\bm p}^2_{\perp} \right ) \psi (\xi, {\bm {\mathfrak R}}) = 0  \label{Eq9.3a}
\end{eqnarray}
Going to dimensionless variables, $\tau = 2\lambda \xi,\, {\bm x}_{\perp} = 2\lambda  {\bm {\mathfrak R}}$, we arrive, again, at Eq.(\ref{Sh.5}).

Then, writing down such normalized solutions to Eq.(\ref{Sh.5}) as a 2D Gaussian wave-packet with the OAM,
\begin{widetext}
\begin{eqnarray}
& \displaystyle \psi (\tau, {\bm x}_{\perp}) = \frac{R}{8 \sqrt{2}\ \sigma} \left ((2\sigma)^{-2} + i\tau\right )^{-3/2} \exp \left\{\frac{i}{2} {\bm \kappa} {\bm x}_{\perp, 0} - \left (\frac{\kappa}{2\sigma}\right )^2 - \frac{{\bm R}^2}{8 ((2\sigma)^{-2} + i\tau)} + i\ell \varphi \right\} \times \cr & \displaystyle \times \left (I_{\frac{1}{2} (\ell - 1)} \left (\frac{{\bm R}^2}{8 ((2\sigma)^{-2} + i\tau)}\right ) - I_{\frac{1}{2} (\ell + 1)} \left (\frac{{\bm R}^2}{8 ((2\sigma)^{-2} + i\tau)}\right )\right ),\cr & \displaystyle {\bm R} = {\bm x}_{\perp} - {\bm x}_{\perp, 0} - i \frac{{\bm \kappa}}{2\sigma^2} \equiv R \{\cos \varphi, \sin \varphi\}\label{Eq9.4}
\end{eqnarray}
\end{widetext}
or an Airy wave-packet,
\begin{widetext}
\begin{eqnarray}
& \displaystyle \psi (\tau, {\bm x}_{\perp}) = \frac{\sqrt{2\pi}}{b_x b_y \sigma} \exp\left\{\frac{i}{2} {\bm \kappa} {\bm x}_{\perp, 0} - \left (\frac{\kappa}{2\sigma}\right )^2 - \frac{1}{3} \left (\frac{1}{b_x^6} + \frac{1}{b_y^6}\right ) \left ((2\sigma)^{-2} + i\tau\right )^3 \right\} \times \cr & \displaystyle \times \text{Ai} \left (b_x^{-1} \left (R_x + \frac{((2\sigma)^{-2} + i\tau)^2}{b_x^3}\right )\right ) \text{Ai} \left (b_y^{-1} \left (R_y + \frac{((2\sigma)^{-2} + i\tau)^2}{b_y^3}\right )\right )\label{Eq9.5}
\end{eqnarray}
\end{widetext}
we arrive at the corresponding wave-packet states of a boson in external field of a plane electromagnetic wave, which are exact solutions of the Klein-Gordon equation.

We would like to emphasize that, in contrast to the external magnetic field case (see e.g., \cite{B-G}), corresponding free-particle states are obtained from here just by putting $A \rightarrow 0$. At the same time, since our states are not localized in longitudinal direction, we imply the adiabatic switching-on and -off of the wave on $t, z \rightarrow \pm \infty$. It means that generically such non-Volkov states stay well-normalized one-particle ones obeying corresponding completeness relations and being non-orthogonal in transverse plane. Though the completeness feature may not be obvious from the afore-given procedure, one can obtain the very same solutions by expanding an arbitrary state of a boson in external field of a wave, $|\psi\rangle$, over the Volkov ``plane-wave'' states, $|{\bm q}\rangle$ , which are known to be orthonormal. This expansion would be analogous in a sense to the ordinary Fourier transform used in the previous sections and could always be reverted. Then by choosing an appropriate physical model for the overlap $\langle {\bm q}|\psi \rangle$, one can obtain the same states, Eqs. (\ref{Eq9.4}), (\ref{Eq9.5}).










\section{Conclusion}

Progress in experimental creation of the electrons having non-planar wave fronts and such new degrees of freedom as, for instance, OAM requires adequate theoretical description of these novel quantum states. In this paper, we presented and analyzed a family of quasi-classical Gaussian wave-packets of massive bosons, which are exact solutions of the Klein-Gordon equation thanks to the null-plane-variables formalism. We compared these exact solutions with the approximate (paraxial) ones and indicated a link between them. Depending upon a choice of the general complex phase in momentum representation, such wave packets can carry OAM and may have finite quantum uncertainty of the latter or can represent well-normalized Airy one-particle states with the zero effective value of the OAM but, again, with its finite uncertainty, which is determined by the packet's parameters. Depending on these parameters, the states obtained can resemble either the OAM-less Gaussian packets (that is the squeezed partially coherent states) or the ``pure Bessel'' and the ``pure Airy'' beams. 

Such features of the calculated quantum OAM-uncertainty as its finiteness and its dual, intrinsic and extrinsic, nature for the Airy states make these wave-packets potentially useful for purposes of quantum entanglement of the electrons and other massive particles in their OAMs, by analogy with the quantum optics. It is easy to show that exactly as in the optical case, a non-vanishing OAM-uncertainty is needed for the final particles in some reaction to be OAM-entangled. Detailed calculations of such a scattering will be presented elsewhere. We believe that the corresponding experiments, similar to those described in Ref.\cite{Ivanov_PRA}, could in principle be carried out with the vortex- and Airy-electrons available at the moment or even with the hypothetical Airy-Bessel electrons.

As also demonstrated, these quantum states exist in external field of a plane electromagnetic wave. This means that all the quantum numbers of vortex (or Airy-) electrons entering the wave would remain (almost) the same after electrons' leaving the wave (neglecting the radiation losses). The last fact, in particular, allows one to propose schemes for acceleration of the vortex- and Airy electrons up to the MeV energies, similar to those we discussed earlier in Ref.\cite{me}.

Finally, generalizations of the wave-packets described in this paper to the fermionic case could be easily obtained by multiplying corresponding wave-functions in momentum representation by a bispinor $u ({\bm p})$ obeying Dirac equation.

\begin{acknowledgments}
I am grateful to A.\,Di~Piazza, I.~Ivanov, C.\,H.~Keitel, and O.~Skoromnik for fruitful discussions and also to the Alexander von Humboldt Foundation for support.
\end{acknowledgments}

\end{document}